\documentclass[12pt,a4paper]{article}
\usepackage{amsmath}
\usepackage{amsfonts}
\usepackage{amssymb}
\usepackage[dvips]{graphicx}
\usepackage{natbib}

\usepackage{ulem}
\bibpunct{(}{)}{;}{a}{}{,}

\usepackage{setspace}
\usepackage{epsfig}   

\usepackage{color}

\newcommand{\ds}{\displaystyle}

\author{Abhinav Singh\\
Department of Biology, LMU Munich\\
\texttt{singh@bio.lmu.de}\\ 
\and
\hskip  -1.6cm 
Hao Wang\\ 
Department of Mathematical and Statistical Sciences, University of Alberta\\
\hskip  -1.5cm 
\texttt{hao8@ualberta.ca}\\
\and
Wendy Morrison\\
School of Biology, Georgia Institute of Technology\\
\texttt{wendymorrison@gatech.edu}\\
\and 
Howard Weiss\\
School of Mathematics, Georgia Institute of Technology\\
\texttt{weiss@math.gatech.edu}}
\date{}

\title{Modeling Fish Biomass Structure at Near Pristine Coral Reefs and Degradation by Fishing}
\begin{document}

\maketitle
\newpage \abstract{Until recently, the only examples of inverted biomass pyramids have been in freshwater and marine planktonic communities. In 2002 and 2008 investigators documented inverted biomass pyramids for nearly pristine coral reef ecosystems within the NW Hawaiian islands and the Line Islands, where apex predator abundance comprises up to 85\% of the fish biomass. We build a new refuge based predator-prey model to study the fish  biomass structure at coral reefs and investigate the effect of fishing on biomass pyramids. Utilizing realistic life history parameters of coral reef fish, our model exhibits  a stable inverted biomass pyramid.  Since the predators and prey are not well mixed, our model does not incorporate homogeneous mixing and the inverted biomass pyramid is a consequence of the refuge. Understanding predator-prey dynamics in nearly pristine conditions  provides a more realistic historical framework for comparison with fished reefs. Finally, we show that fishing transforms the inverted biomass pyramid to be bottom heavy. }
 
\section{Introduction}

An inverted biomass pyramid has increasing biomass along trophic levels~\citep{odu71}. Inverted biomass pyramids in ecology are highly counterintuitive and appear to be exceedingly rare. Inverted biomass pyramids have only been observed in aquatic planktonic communities~\citep{odu71, buc96,  gas97, del99, mou06}. Odum~\citep{odu71} hypothesized that the high turn-over rate and the metabolism of phytoplankton can produce inverted biomass pyramids. Other hypotheses include the low turn-over rate of predators~\citep{cho90, del99} and the influx of organic matter which act as food for heterotrophic predators~\citep{del99}.

Recently, inverted biomass pyramids have also been observed at coral reefs where up to 85\% of the fish biomass was composed of apex predators~\citep{fri02, san08}. Historical observations suggest that this high abundance of predators was common~\citep{san08}  and such reefs can be considered \lq nearly pristine\rq\ ~\citep{knowlton2008sbl}, and thus provide a baseline for studying natural reefs. The coral cover at these pristine reefs is far more extensive and healthier than at conventional reefs, and these reefs seem to be either resistant or resilient to ocean warming and rising acidity~\citep{knowlton2008sbl, san08}.

The high predator biomass at these \lq nearly pristine\rq\ reefs is in sharp contrast to most reefs, where the prey biomass substantially dominates the total fish biomass~\citep{san08}. The mechanisms causing inverted biomass pyramids in planktonic communities and coral reefs are clearly different. The former relies upon homogeneous mixing which is clearly not present in coral reefs with many holes (as the refuge) for prey to hide.

Some ecologists believe that refuges provide a general mechanism for interpreting ecological patterns~\citep{haw93}. Previous experimental and theoretical studies of prey refuges have demonstrated how refuges increase the abundance of prey and add stability to the system~\citep{huf58, berryman2006refuge}. Few studies have analyzed the impact of refuges on predator abundances (see~\citep{per95}  for an analysis of how refuges impact predator growth), and none have addressed how refuges affect predator to prey biomass ratios. We study the influence of refuges on prey growth, predator feeding behavior, and predator and prey biomass in this manuscript.

Investigations into the importance of shelter for coral reef fish abundance have found mixed results. Robertson~\citep{robertson1981availibility} monitored the abundance of reef fish after physically removing half of the patch reef. He found that the abundance of a small herbivorous damselfish on the remaining reef increased by up to 1.63 times the original density, suggesting that shelter was not limiting. Conversely, Shulman~\citep{shu84} found that the presence of shelter increases the recruitment and survival of small herbivorous fish. Similarly, Holbrook and Schmitt~\citep{holbrook2002css} used infra-red photography at night to document the predation of small reef fish that were located near or outside the coral shelter. Extremely high mortality of coral reef fish (up to 60\%) has been documented during settlement~\citep{doh85} and during the days directly after settlement~\citep{almany2006pge}. This is followed by a reduction in mortality as time passes~\citep{doh85}. This suggests that mortality during and directly after settlement may be a population bottleneck for reef species~\citep{doherty2004hmd}. Doherty and Sale~\citep{doh85} showed that providing a cage during this time decreased predation on sedentary reef species, though they were unable to quantify natural mortality due to confounding factors.

Mathematical modeling can provide insights into some of the fundamental open questions about the biomass structure at coral reefs. Classical predator-prey models, including Holling type models, assume that predators and prey are well mixed, i.e., all prey are accessible to the predators. This is not the case at coral reefs where small fish find \lq refuge\rq\ from predators in coral holes where large predators cannot enter~\citep{hix93}. Thus, a population model assuming homogeneous mixing between predators and prey is not appropriate to study the fish biomass structure at coral reefs. An appropriate model must include a prey refuge and its associated functional response of predation due to the refuge.

  In our recently published work Wang et al. \ \citep{wang2009mib}, we used mathematical modeling to investigate the theoretical conditions necessary for the creation of inverted biomass pyramids, but we did not test the ideas with actual predator and prey life history parameters. In addition, we developed a family of predator-prey models (RPP model) that explicitly incorporate a \lq prey refuge\rq, where the refuge size influences predator hunting patterns (predation response). We showed that refuges provide a new general mechanism in ecology to create an inverted biomass pyramid that does not require mass action interactions between predators and prey.

This study is an extension as well as a modification of Wang et al. ~\citep{wang2009mib}'s work by modeling the coral reef inverted biomass pyramids with realistic life-history parameters specific for coral reef fish. Our new refuge-based predator-prey model exhibits a stable inverted biomass pyramid and thus provides a mechanism to explain the recently observed inverted biomass pyramid at nearly pristine reefs. We end this paper by using the parameterized model to investigate the impact of fishing on the biomass ratio.

\section{\label{derivation}Derivation of the Model}
Guided by field observations at pristine coral reefs, we derive a model for the biomass of coral reef fishes using a pair of differential equations. Following Sandin et al.~\citep{san08}, we classify reef fishes as prey or predators. We model prey when they are large enough to be visualized by the divers on the survey and are a possible source of food for the apex predators (i.e. past the high mortality experienced during recruitment). We include herbivores, and planktivores within our prey categories, and include the top predators within our predator category. We currently have not incorporated the carnivores (i.e. small predators) into the model as they consume mainly small invertebrates (Sandin et al. 2008) and thus have minimal impact on the abundance of prey fishes.

Prey fish eat plankton and algae and hide from predators in coral holes~\citep{pala2007rtl,san08, hix93, caley1996ras}. We assume that prey biomass grows logistically and (per capita predator) predation rate depends on prey biomass and availability of coral holes to hide. Predators grow by eating prey fish and die a natural death at pristine reefs. Prey fish find \lq refuge\rq\ in coral holes and rarely venture out of the holes at Kingman~\citep{pala2007rtl}. Therefore, the availability of hiding space for prey in coral holes affects predator hunting patterns and thus the biomass pyramid. We define the \lq refuge size\rq\ as the maximum prey biomass which can sustainably hide in coral holes, i.e. the coral-specific prey carrying capacity in presence of predators~\citep{daily1992psa}. We distinguish the refuge size from the prey carrying capacity in absence of predators (K); the prey will not be forced to stay inside the holes when the predators are absent and the reef can support a much greater prey biomass. We assume that the refuge size is an increasing function of coral cover at pristine reefs. The equations describing such a community are

\begin{align}
\label{preyeq}\frac{dx}{dt} &=a(r)x\left(1-\frac{x}{K}\right)  - bf(x,r)y,\\
\label{predeq}\frac{dy}{dt} &=cbf(x,r)y - dy.\\\notag\\
\notag x&: \text{prey biomass density} ~\mathrm{(kg/m^2)},\\
\notag y&: \text{predator biomass density} ~\mathrm{(kg/m^2)},\\
\notag a(r)&: \text{prey growth rate} ~\mathrm{(/day)},\\
\notag b&: \text{maximum predation rate: maximum prey biomass},\\
\notag &~~ \text{  hunted per kg of predator biomass (/day)},\\
\notag K&: \text{prey carrying capacity in absence of predators}~\mathrm{(kg/m^2)},\\
\notag r&: \text{refuge size} ~\mathrm{(kg/m^2)},\\
\notag f(x,r)&: \text{predation response},\\
\notag c&: \text{biomass conversion efficiency},\\
\notag d&: \text{predator death rate (/day)}.
\end{align}

The estimated annual mortality rates of small reef fish can be as high as 5-6 ~\citep{kritzer2002ssm, wilson2004gma}, suggesting that in the absence of predation, prey fish can double in 2-3 months.  Therefore for our model, prey growth rate varies between 0.003 and 0.007, which is equivalent to prey doubling every 7 and 3 months, respectively. Predator death rate (d=0.0005/day) was estimated using the equation:  d = -ln (0.01)/longevity~\citep{mollet2002cpd}, with the estimated longevity for grey reef shark of 25 years~\citep{froese2008fv}.  We set prey carrying capacity at  K=2 $\mathrm{kg/m^2}$, roughly seven times the maximum prey biomass measured at Kingman reef~\citep{demartini2008dfa}.  We set the biomass conversion efficiency (c)  to 0.15, a reasonable estimate given that conversion efficiencies are higher in marine versus terrestrial environments~\citep{rskrichardstephenkentbarnes}. Predation rates of 12\% predator body weight per day have been documented for smaller sedentary predators~\citep{sweatman1984fsp}, suggesting that rates for active predators would be higher.  We therefore set the maximum predation, b=0.24/day.

Wang et al.~\citep{wang2009mib} developed the family of refuge-modulated predator prey models (RPP Type I, II and III) to explicitly include the multiple effects of refuge on the feeding behavior of predators. The effects of a refuge can be included by the generalized predation response function 
 \begin{align}
	f(x,r)&=\frac{1}{1+e^{-\xi[x-(2-i)r]}},\\
	\notag i&=1, 2 ~\text{and}~ 3,
\end{align} \\

The choice of i depends on the environment under consideration.  Adding a refuge to the ecosystem could conceviably either decrease the prey available to the predators ($i=1$; RPP is Type I), have no impact on the number of prey available ($i=2$; RPP is Type II), or increase the number of prey available to predators ($i=3$, RPP is Type III).
The predation response function $f(x,r)$ at coral reefs should have the following properties. It should be a monotonically increasing function of prey biomass. When the prey  biomass is less than the refuge size, it should be small.  When prey biomass  approaches refuge  size, it should rapidly increase and as prey biomass greatly exceeds the refuge size, the predators become satiated and the response function approaches a constant; thus forming an S shaped curve. We believe the predation function from RPP Type I
\begin{equation}
f(x,r)=\frac{1}{1+ e^{-10(x-r)}}\\
\end{equation}
is the simplest function having these properties.\\

\begin{figure}[htbp]
\center\includegraphics[width=3in]{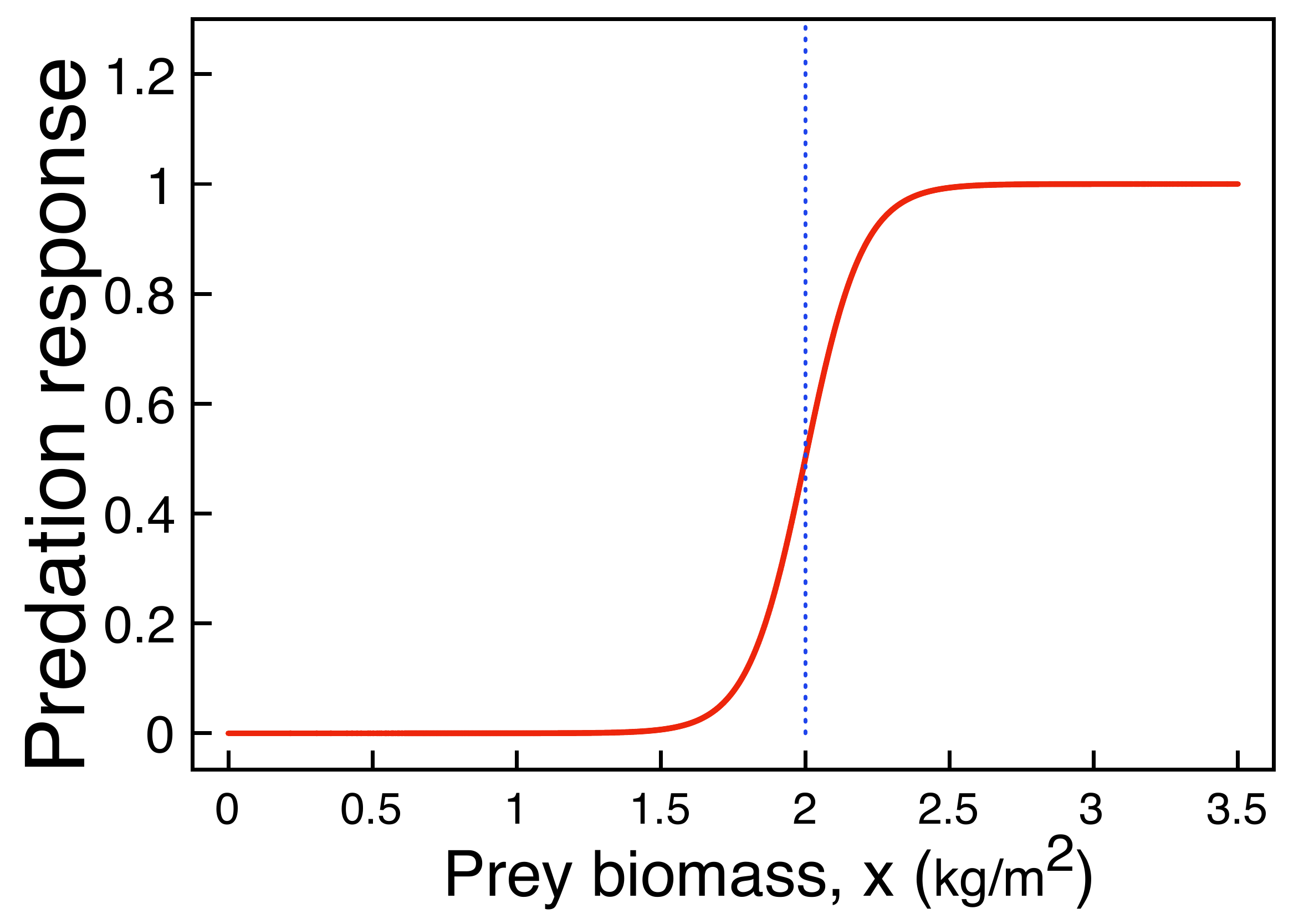}
\caption{\label{predation} Predation function $f(x,r)$ vs biomass of prey for fixed refuge size~$r$=$2~ \mathrm{kg/m^2}$.}
\end{figure}

Figure~\ref{predation} is a plot of $f(x,r)$ for fixed refuge size of $2~ \mathrm{kg/m^2}$.\\

\subsection{Influence of refuge on prey productivity}
It has been suggested that juvenile fish  mortality during and directly after settlement can create a population bottleneck~\citep{doherty2004hmd}. Research shows that an increase in the refuge size can increase the survival rate of juveniles~\citep{doh85, shu84}  and may increase the prey growth rate.  Since we have defined prey abundance as the number of prey that have survived to a size where they are visually detectable and viable food for the top predators, increasing the shelter available to recruits will increase the number of fish that become available prey. This idea is similar to the idea of recruitment within fisheries science where fish are considered recruits when they have reached a size where they can be captured by the fishery. For this reason, we include in our model a variable prey growth rate dependent on the refuge size, i.e. $a(r)$. We model $a(r)$ as a sigmoid curve where at low refuge cover, there is low survival of recruits, with survival increasing to some upper level where saturation of refuges for recruits results in an asymptote;  we use the function
\begin{equation}
 a(r)=0.003 + \left( \frac{0.004r^{12}}{0.1 + r^{12}}\right).\\
\end{equation}
\\

\begin{figure}[htbp] 
   \centering
   \includegraphics[width=3in]{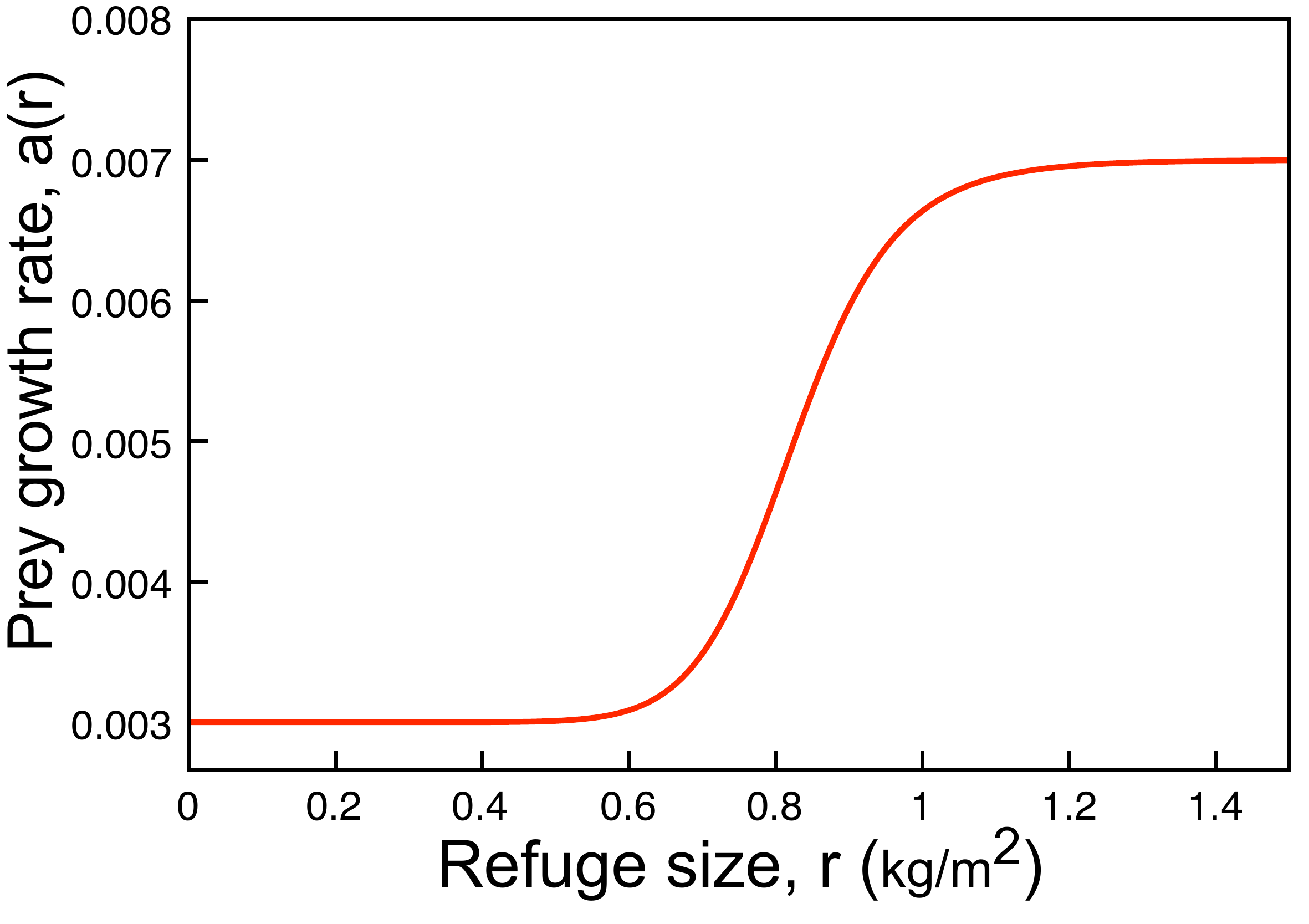} 
   \caption{Prey growth rate, $a(r)$ is a function of the refuge size.}
   \label{fig:varpreygrowth}
\end{figure}

We plot the refuge-dependent prey growth rate $a(r)$ in Figure~\ref{fig:varpreygrowth}.

The following equations describe the complete model:
\begin{align}
\label{preyeq2}\frac{dx}{dt} &=\left(0.003 +  \frac{0.004r^{12}}{0.1 + r^{12}}\right)x\left(1-\frac{x}{K}\right)  - \frac{by}{1+ e^{-\ds10(x-r)}},\\
\label{predeq2}\frac{dy}{dt} &=c\frac{by}{1+ e^{-\ds10(x-r)}} - dy.\\\notag
\end{align}

\section{Results}
The system of differential equations has three equilibrium points. The unstable equilibrium point, $x=0,~ y=0$ corresponds to a reef with no fish.  The equilibrium point $x=K,~y=0$ corresponds to  the absence of predators and is rarely seen in reefs. The third and the most interesting equilibrium point, which we call the interior equilibrium point is
\begin{eqnarray}
x^{*}(r)&=&r- \frac{1}{10} \ln \left(\frac{bc}{d}-1\right),\\
y^{*}(r)&=&\frac{a(r)c}{d}x^{*}\left(1-\frac{x^{*}}{K}\right).
\end{eqnarray}
This equilibrium point is locally attractive for the refuge size between 0.65-0.9$~\mathrm{kg/m^2}$. The predator-prey biomass ratio at the third equilibrium point is
\begin{equation}
\label{brationofish}\frac{y^{*}(r)}{x^{*}(r)}= \frac{a(r)c}{d}\left(1 +  \frac{1}{10K}\ln \left (\frac{bc}{d}-1 \right ) - \frac{r}{K}\right).\\
\end{equation}

\begin{figure}[htbp]
\begin{center}

\includegraphics[width=3in]{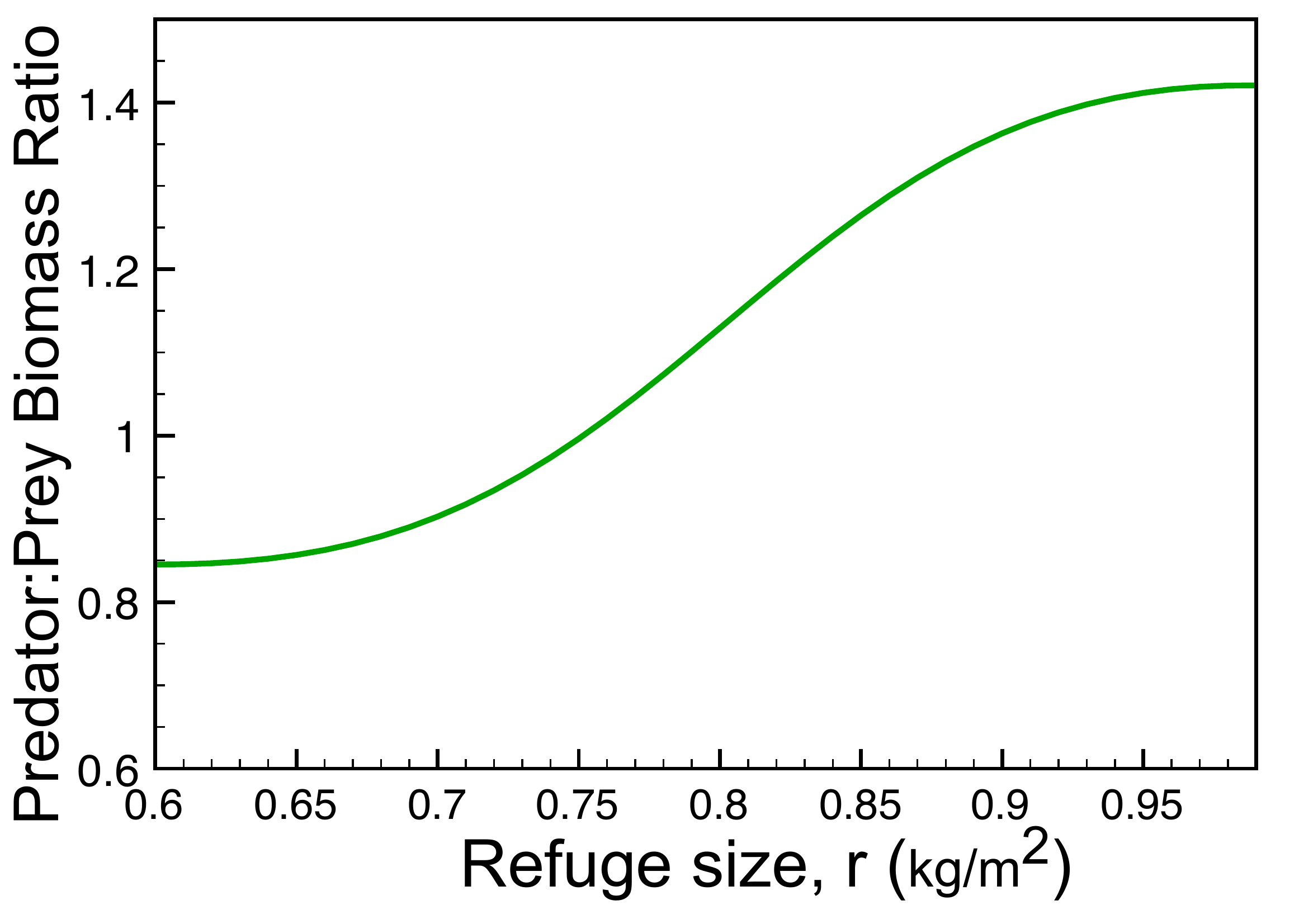}
\caption{\label{ratioa}The biomass pyramid is inverted and the predator:prey biomass ratio is an increasing function of refuge size.}
\end{center}
\end{figure}

Figure~\ref{ratioa} illustrates the dependence of the predator-prey biomass ratio on the refuge size. The predator-prey biomass ratio is now an increasing function of refuge size, a prediction supported by data from Kingman and Palmyra. The coral cover at Kingman is more extensive than Palmyra: predators constitute 85\% of the fish biomass at Kingman while they constitute only 66\% of the fish biomass at Palmyra~\citep{san08}.\\

\section{Effects of Fishing}
\label{fishing}
It is believed that fishing can dramatically change the biomass ratio; the fish biomass pyramid becomes bottom heavy  at reefs with fishing \citep{san08, warne2008aue}. We add fishing to our model and show that sufficiently high fishing pressure will destroy the inverted pyramid. Destruction of the inverted pyramid in presence of predator fishing is direct, but we show that prey fishing alone will also destroy the inverted biomass pyramid.\\

\begin{figure}[htbp]
\begin{tabular}{cc}
\includegraphics[width=3in]{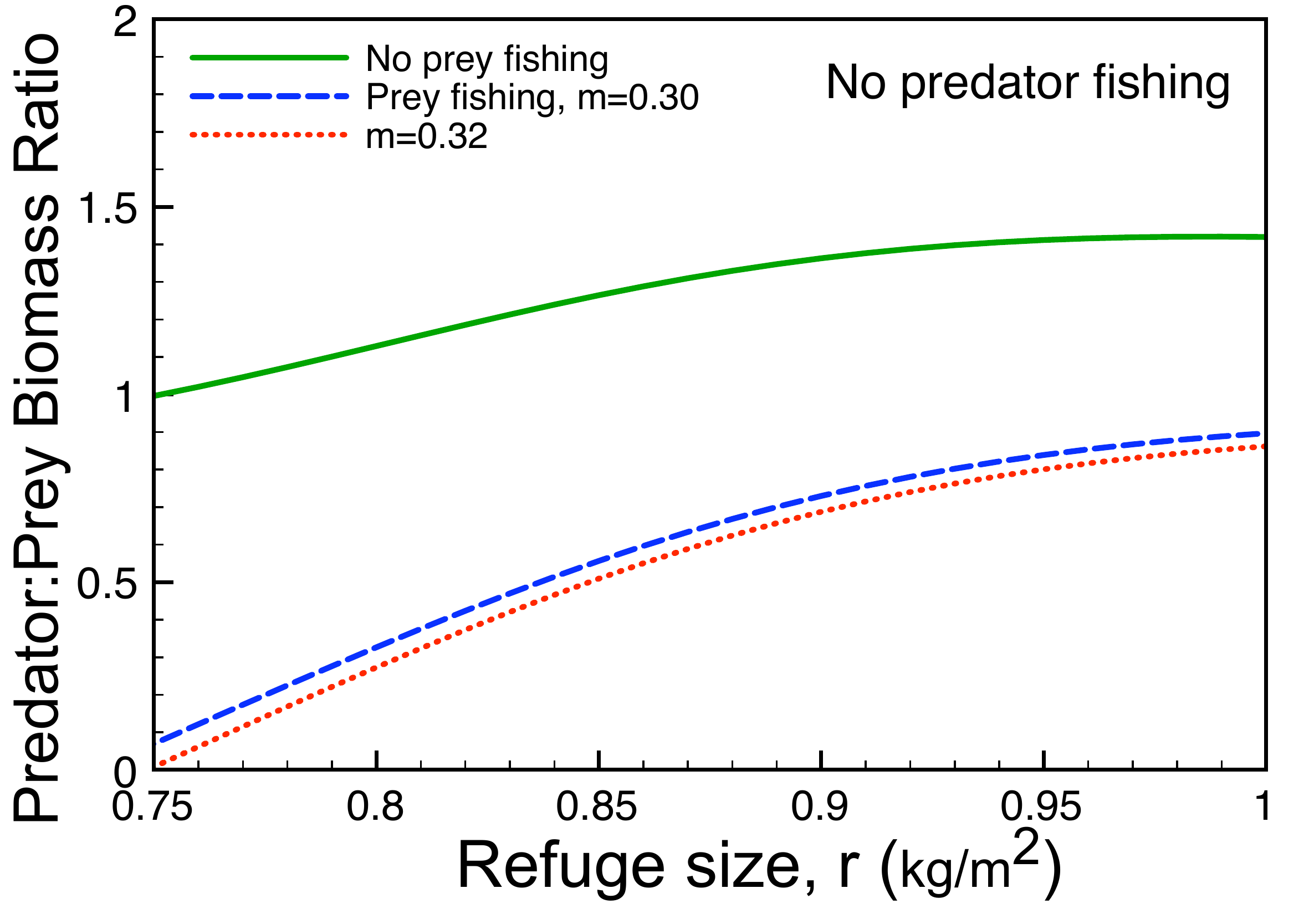}&
\includegraphics[width=3in]{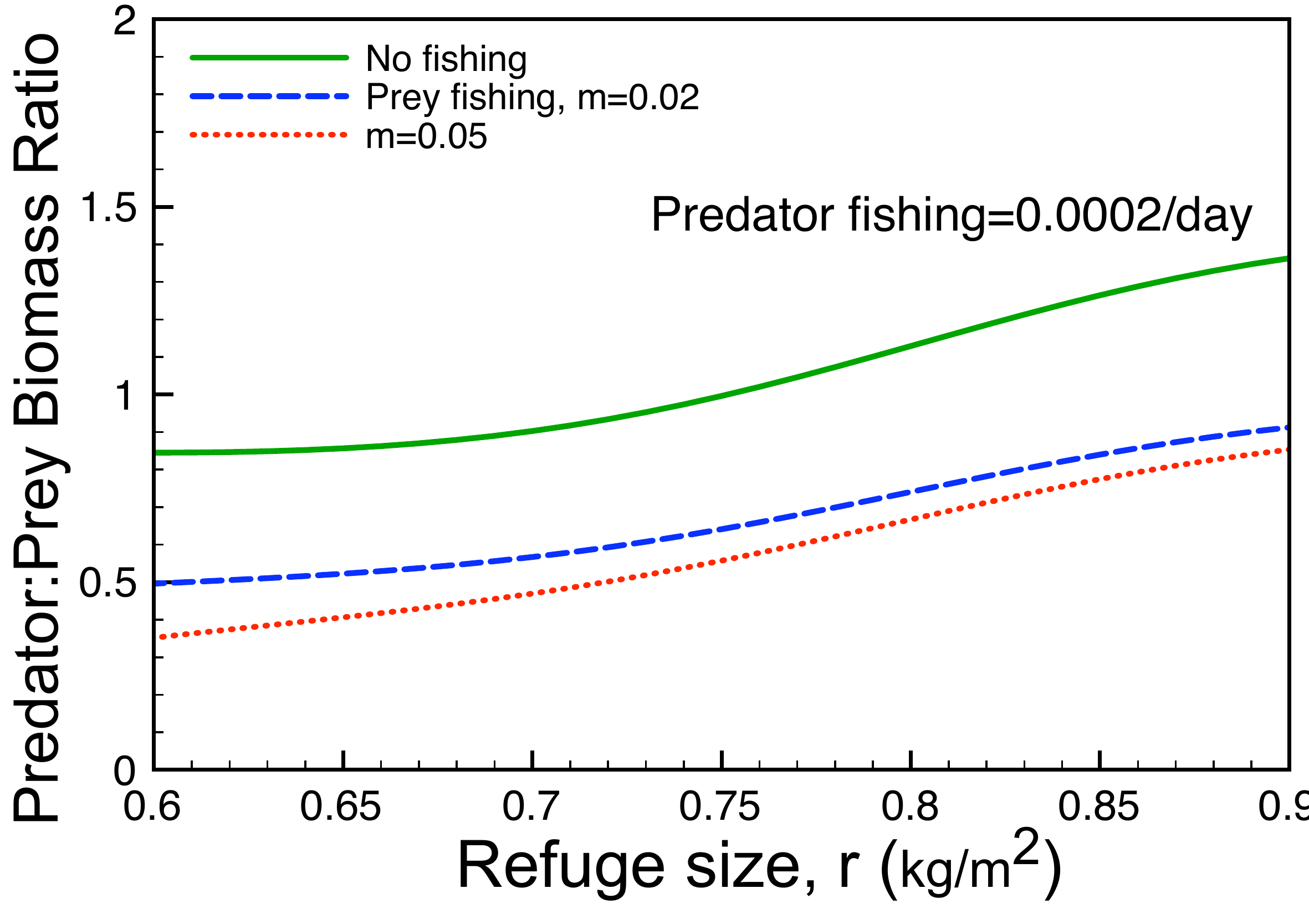}\\
(a)&(b)
\end{tabular}
\caption{\label{ratiowfishing}Predator-prey biomass ratio as a function of refuge size with different prey fishing effort(e). Parameters: $K=2.0, b=0.24, d=0.0005$, predator fishing rate: (a) $l=0$; (b) $l=0.0002$.}
\end{figure}

\noindent As an illustrative example, we assume that predator fishing rate is proportional to the predator biomass and prey fishing is similar to predator hunting. We understand that this is not the only form of prey fishing and thus we further show that our results are qualitatively robust to changes in forms of prey fishing. The model equations incorporating fishing are
\begin{align}
\frac{dx}{dt} &= a(r)x\left(1-\frac{x}{K}\right)  - b\frac{y}{1+ e^{-\ds10(x-r)}} -b\frac{m}{1+ e^{-\ds10(x-r)}}\\
\frac{dy}{dt} &= cb\frac{y}{1+ e^{-\ds 10(x-r)}} - dy -ly,\\
\notag m &: \text{Prey fishing effort (/day)},\\
\notag l &: \text{Predator fishing effort (/day)}.\\
\end{align}
\\The prey and predator biomass at the interior equilibrium point are
\begin{eqnarray}
\label{preyeqfish}\tilde{x}(r,l)&=&r- \frac{1}{10} \ln \left(\frac{bc}{(d+l)}-1\right),\\
\tilde{y}(r,l)&=&\frac{a(r)c}{(d+l)}\tilde{x}(r,l)\left(1-\frac{\tilde{x}(r,l)}{K}\right) -m.
\end{eqnarray}
\noindent The new predator-prey biomass ratio at the interior equilibrium point is 
\begin{align}
\label{bratiofishrefuge}\frac{\tilde{y}(r,l)}{\tilde{x}(r,l)}&= \frac{a(r)c}{d+l}\left(1- \frac{\tilde{x}(r,l)}{K}\right) - \frac{m}{\tilde{x}(r,l)},\\ \quad \text{with} \quad \tilde{x}(r,l)&=r- \frac{1}{10} \ln \left(\frac{bc}{(d+l)}-1\right).
\end{align}
We plot the predator-prey biomass ratio for various refuge sizes and fishing rates in Figure~\ref{ratiowfishing}.

We now deduce the effect of fishing on the predator-prey biomass ratio by  inspecting Figure~\ref{ratiowfishing} and comparing equation ~\eqref{bratiofishrefuge} with equation ~\eqref{brationofish}: \textit{the predator-prey biomass ratio is a decreasing function of fishing pressure and the biomass pyramid becomes bottom heavy (ratio less than unity) at conventional coral reefs that experience high fishing pressure}. Figure~\ref{ratiowfishing}(a)  shows that the biomass ratio decreases even with prey fishing only and this makes the pyramid  bottom heavy.

Our results are independent of the form of prey fishing. Let $p(x)$ be the general prey fishing rate. The modified equations are
\begin{eqnarray}
\frac{dx}{dt} &=& a(r)x\left(1-\frac{x}{K}\right)  - b\frac{y}{1+ e^{-\ds 10(x-r)}}- p(x)\\
\frac{dy}{dt} &=& cb\frac{y}{1+ e^{-\ds10(x-r)}} - dy - ly.
\end{eqnarray}
\\The predator-prey biomass ratio at the interior equilibrium point is 
\begin{align}
\label{bratiowgenfish}\frac{\tilde{y}(r,l)}{\tilde{x}(r,l)} &= \frac{a(r)c}{d+l}\left(1 +  \frac{1}{10K}\ln \left (\frac{bc}{d}-1 \right ) - \frac{r}{K}\right) - \frac{c}{d+l}\frac{p(\tilde{x})}{\tilde{x}},
\end{align}
the biomass ratio at the fished reef ($\tilde{y}(r,l)/\tilde{x}(r,l)$) is lesser than the biomass ratio at a reef without fishing ($y^{*}(r)/x^{*}(r)$)
\begin{align}
\frac{\tilde{y}(r,l)}{\tilde{x}(r,l)} &\leq \frac{a(r)c}{d}\left(1 +  \frac{1}{10K}\ln \left (\frac{bc}{d}-1 \right ) - \frac{r}{K} \right )=\frac{y^{*}(r)}{x^{*}(r)}.
\end{align}\\

As a result of fishing, the predator-prey biomass ratio is  less than the biomass ratio at reefs without fishing. This result is robust under different forms of fishing.\\

As another example of prey fishing, if the prey fishing rate is proportional to prey biomass, $p(x)= vx$, the predator-prey biomass ratio 
\begin{equation}
\frac{\tilde{y}(r,l)}{\tilde{x}(r,l)}= \frac{a(r)c}{d+l}\left(1+  \frac{1}{10K}\ln \left (\frac{bc}{d}-1 \right ) - \frac{r}{K}\right) - \frac{c}{d+l}v.
\end{equation}
This is less than the biomass ratio for the model without fishing in Equation~\eqref{brationofish} and high fishing pressure will destroy the inverted biomass pyramid. \\

\section{Discussion}

In this manuscript, we model the fish biomass structure in near pristine coral reef ecosystems and our model displays a stable inverted biomass pyramid. We show how the presence of refuge can influence the inverted biomass pyramid through the modification of prey growth rate and predator response function. Our model confirms previous suggestions that high prey growth rate and low predator growth rate are necessary for inverted biomass pyramids~\citep{odu71, del99, cho90}. Both conditions are satisfied at \lq nearly pristine\rq~ reefs where apex predators such as sharks can live up to 20 years and reproduce rarely~\citep{smith1998irp} and smaller prey fish can reproduce at least three times a year~\citep{srinivasan2006eba}. In addition, we show that sufficiently high fishing pressure will destroy the inverted biomass pyramid.

By incorporating realistic parameter values, we show that inverted biomass pyramids on reefs are possible. Coral holes are essential to our model as prey fish at pristine reefs take \lq refuge\rq~ in coral holes from predators and were rarely observed to leave the holes~\citep{san08}. Prey fish also practice\lq hot-bunking\rq, i.e. if one prey fish left a coral hole, another immediately occupied that hole~\citep{pala2007rtl}. Our model assumes that the refuge size influences prey growth rate. The protection provided to juveniles by the coral cover ends up boosting the overall supply of prey fish by increasing prey growth rate a(r).  Alternatively, this same concept could be incorporated into the model through the RPP Type III equation~\citep{wang2009mib}, but for the parameter values implemented here, leads to an unrealistic unstable biomass pyramid. If we assume that prey survival to adult size is dependent on the cover of coral reef, we find that the predator-prey biomass ratio is an increasing function of refuge size. This relationship is supported by data from~\citep{san08} comparing Palmyra and Kingman.

The predator and prey life-history estimates utilized for this paper are at the extremes of those measured in the field. For example, the  prey growth rate variation from 0.003 to 0.007 applies to small planktivorous fish. Larger herbivores (i.e. parrotfish) have much lower growth rate estimates (.0013/day; Fishbase). However, all available parameter estimates are from the highly impacted reefs with low predator abundances. No estimates exist for life history parameters of coral reef fish at  any of the locations with the inverted biomass structure.

When the fishing pressure is sufficiently strong, the inverted biomass pyramid disappears (see Figure~\ref{ratiowfishing}). This is consistent with field observations where reefs with fishing exhibit a non-inverted bottom heavy pyramid~\citep{san08}. Our model shows that the biomass ratio decreases when either predator or prey fishing or a combination of both takes place.  Further computations, which we do not present,  show that prey fishing alone can have the same effect.

\section{Appendix}
\subsection{Local Stability of equilibrium points\\}
The equations governing the dynamics of predator and prey biomass are described by
\begin{align}
\notag\frac{dx}{dt} &=a(r)x\left(1-\frac{x}{K}\right)  - bf(x,r)y,\\
\notag \frac{dy}{dt} &=cbf(x,r)y - dy.\\\notag
\end{align}
The equilibrium points are (0, 0), (K, 0) and ($x^{*}, y^{*}).$
\begin{align} 
   x^{*}&=r- \frac{1}{10} \ln\left(\frac{bc}{d}-1\right),\\
   y^{*}&=\frac{a(r)c}{d}x^{*}\left(1-\frac{x^{*}}{K}\right).
\end{align}
We determine the local stability of the equilibrium points by computing the Jacobian at the equilibrium points. The Jacobian 
\\
\begin{equation*}
\ds J =  \left[\begin{array}{cc}
a(r)- 2a(r)\frac{\ds x}{K} - 10by\ds \frac{\ds(\ds e^{-\ds 10(x-r)})}{(1+ e^{-\ds10( x-r)})^2} & -\ds \frac{b}{1+ e^{-\ds 10(x-r)}} \\\\
4bc\ds \frac{y(e^{-\ds 10(x-r)})}{(1+ e^{-\ds10(x-r)})^2} & \ds \frac{bc}{1+ e^{-\ds 10(x-r)}} -d
\end{array}\right]. \\
\end{equation*}

At (0,0)\\
\begin{equation*}
J(0,0) =  \left[ \begin{array}{cc}
a(r) & -\ds \frac{b}{1+ e^{\ds10r}} \\\\
0 & \ds \frac{bc}{1+ e^{\ds 10r}} -d
\end{array}\right]. \\
\end{equation*}
The eigenvalues of the Jacobian are $a(r)$ and $(bc/{(1+ e^{10r})} -d)$. As $a(r) \geq 0$, (0,0) is an unstable equilibrium point~\citep{strogatz1994nda}.\\

At (K,0),
\begin{align*}
J(K,0)& =  \left[\begin{array}{cc}
-a(r)  & -\ds\frac{b}{1+ e^{-\ds10(K-r)}} \\\\
0 & \ds\frac{bc}{1+ e^{-\ds10(K-r)}} -d\\
\end{array}\right]\\
\text{and}~\det(J(K,0))& = -a(r)\left(\frac{bc}{1+ e^{-\ds10(K-r)}} -d\right) < 0 .\\
\end{align*}

As $1+ e^{-\ds 10(K-r)} \leq 2$ and $bc  > 2d $, $\det(J(K,0))  < 0 $
. Therefore, (K,0) is a saddle equilibrium point~\citep{strogatz1994nda}.\\

At $ (x^{*},y^{*})$,\\
\begin{align*}
x^{*}(r)&=r- \frac{1}{10} \ln \left(\frac{bc}{d}-1\right),\\
y^{*}(r)&=\frac{a(r)c}{d}x^{*}\left(1-\frac{x^{*}}{K}\right),\\
J(x^{*},y^{*}) &=  \left[\begin{array}{cc}
a(r)- 2a(r)\frac{\ds x^{*}}{K} - 10by^{*}\ds\frac{e^{-\ds10(x^{*}-r)}}{\ds(1+ e^{-\ds10(x^{*}-r)})^2} & \ds\frac{-b}{1+ e^{-\ds10(x^{*}-r)}} \\\\
10bc \ds \frac{y^{*}e^{-\ds10(x^{*}-r)}}{(1+ e^{-\ds10(x^{*}-r)})^2} & 0
\end{array}\right], \\\\
\end{align*}
\begin{align*}
\det J(x^{*},y^{*})&= \ds\frac{10a(r)cx^{*}(1-x^{*}/K)(\frac{\ds c}{\ds d}-1)}{b(\frac{\ds c}{\ds d})^2},\\\\
\operatorname{Tr}(x^{*},y^{*})& = a(r)- 2a(r)\frac{x^{*}}{K} - 10by^{*}\ds\frac{e^{-\ds 10(x^{*}-r)}}{(1+ e^{-\ds 10(x^{*}-r)})^2}.\\
\end{align*}

The determinant and the trace of the Jacobian are complicated functions of the parameters and equilibrium predator and prey biomass. Computer assisted  analysis shows that $\det J(x^{*},y^{*}) \geq 0 $ and $\operatorname{tr} J(x^{*},y^{*}) \leq 0$ when $ 0.60 \leq r \leq 0.99$. Therefore, $(x^{*},y^{*})$ is an attractive equilibrium point when $ 0.60 \leq r \leq 0.99$.

\subsection{Sensitivity analysis\\}
We determine the sensitivity of the predator:prey biomass ratio to variation in the parameters of the equations ~\eqref{preyeq}, \eqref{predeq} and \eqref{brationofish} by means of a sensitivity index. The normalized forward sensitivity index of a variable to a parameter is the ratio of the relative change in the variable to the relative change in the parameter~\citep{chitnis2008dip}. 
As an example, the sensitivity of the biomass ratio to variation in maximum predation rate (b) is given by 
\[ \gamma^{ratio}_b = \frac{\partial ratio}{\partial b}. \frac{b}{ratio} = \left(\frac{a(r)c}{10Kd}\right)\left(\frac{1}{(bc/d) -1}\right)\frac{c}{d}\frac{b}{ratio}. \]\\

The absolute value and the sign of the sensitivity index both contain useful information. The absolute value measures the sensitivity of the variable to variation in the parameter:  a low absolute value denotes robustness in the value of the variable to variation in the parameter and vice versa. A positive sensitive index for a parameter shows that the variable is an increasing function of the parameter.\\

\noindent Table~\ref{sensitivity} shows the sensitivity index for each parameter and organizes them in decreasing order of influence on the biomass ratio.
 
\begin{table}[htbp]
\begin{center}
\begin{tabular}{cc}
\hline
Parameter & Sensitivity Index\\
\hline
r & 1.55\\
c & 0.61\\
d & -0.61\\
K & 0.11\\
b & 0.05\\
\hline
\end{tabular}
\caption{Sensitivity indices for parameters in equations~\eqref{preyeq}, \eqref{predeq} and \eqref{brationofish}. Baseline value for parameters:( $b=0.24, c=0.15, d=0.0005, K=2.0, r=0.9$, biomass ratio$=1.13$).\label{sensitivity}}
\end{center}
\end{table}%

The predator:prey biomass ratio is most sensitive to variation in the refuge size (r)  and least sensitive to variation in the predation response (b). The signs of the sensitivity indices tell us that the predator:prey biomass ratio is an increasing function of  r (per unit area coral reef refuge size), b (maximum predation rate), c (biomass conversion efficiency) and  K (prey carrying capacity) and a decreasing function of  d (predator death rate). 

\newpage
\singlespacing
\bibliographystyle{coral_reef}

\bibliography{fishrefabrev,refugebibabrev}
\end{document}